\documentclass[12pt]{iopart}

\usepackage{iopams}  
\usepackage{bm}
\usepackage{graphicx}
\usepackage{amssymb}
\usepackage{hyperref}

\def\la{\label}
\def\beq{\begin{equation}}
\def\eeq{\end{equation}}
\def\bea{\begin{eqnarray}}
\def\eea{\end{eqnarray}}
\def\p{\partial}

\begin{document}

\title{Relaxation of nonlinear oscillations in BCS superconductivity}

\author{Razvan Teodorescu}

\address{Physics Department, Columbia University, 538 West 120th Street, Mail Code 5293, New York, NY 10027}
\ead{rteodore@phys.columbia.edu}

\begin{abstract}
The diagonal case of the $sl(2)$ Richardson-Gaudin quantum pairing model \cite{Richardson1,Richardson2,Richardson3,Richardson4,Richardson5,Richardson6,Gaudin76} is known to be solvable as an Abel-Jacobi inversion problem \cite{SOV,Kuznetzov,Kuz1,Kuz2,Kuz3,Kuz4,Kuz5,YAKE04}. This is an isospectral (stationary) solution to a more general integrable hierarchy, in which the full time evolution can be written as isomonodromic deformations. Physically, the more general solution is appropriate when
the single-particle electronic spectrum is subject to external perturbations. The asymptotic behavior of the nonlinear oscillations in the case of elliptic solutions is derived.

\end{abstract}



\section{Introduction}
      \label{sec:intro}

The integrable system described by the pairing hamiltonian introduced by Richardson and Sherman \cite{Richardson1,Richardson2,Richardson3,Richardson4,Richardson5,Richardson6} in the context
of nuclear physics has received revived interest in recent years, after being applied to metallic  superconducting grains 
\cite{Sierra} and cold fermionic systems 
\cite{Levitov,YAKE04}. The model is intimately related \cite{CRS97,ALO21,DES01} to a class of integrable systems 
generally referred to as Gaudin magnets \cite{Gaudin76}. These systems have been 
studied both at quantum and classical level \cite{Dukelsky,AFS,Samtleben,Hikami,Gawedzki1,Gawedzki2,Ortiz,Korotkin,Harnad,Dilorenzo}, in the elliptic case as well as trigonometric and rational degenerations, using various methods  from integrable vertex models to singular limits of Chern-Simons theory. 

In \cite{Levitov}, an interesting regime of the pairing problem was considered, which may be relevant to recent experiments with cold fermionic
gases exhibiting the paired BCS state \cite{Fermi1,Fermi2,Fermi3,Fermi4,Fermi5,Fermi6,Fermi7,Fermi8}. It was shown 
that for such systems, the time scales of the order parameter $\tau_{\Delta}
\sim |\Delta|^{-1}$, and the quasiparticle energy relaxation time $\tau_{\epsilon}$ are both much larger than typical time for switching
on the pairing interaction $\tau_0$, essentially given by the variation of external parameters, such as detuning from the Feshbach resonance. It was 
argued that in this regime, for times $t \ll \tau_{\epsilon}$, the dynamics
of the system is given by non-linear, non-dissipative equations describing the coherent BCS fluctuations for the system out of equilibrium. In this limit, the system is integrable, and features non-perturbative behavior, such as soliton-type solutions.

In the mean-field limit, such non-trivial solutions describing the collective mode of the Anderson spins \cite{Anderson} were derived in \cite{Levitov}, for a two-level effective system. This work was generalized \cite{YAKE04} in algebro-geometrical terms. 

In \cite{Gurarie,Leggett,YTA05}, the long-time behavior of the solution has been considered, under various conditions. An issue not addressed so far is the relaxation of the nonlinear oscillatory solution induced by perturbations of the spectral curve, physically justified by coupling to the environment. Several possible kinds of perturbations may be considered, which may lead to different 
types of relaxation.
 
In this paper, we consider the effect of fluctuations of single-particle energy levels, which amount to slow deformations of the Liouville tori, and can be described by hydrodynamic-type equations in phase space \cite{Krichever,K1}. These equations  describe the evolution of the moduli for the complex curve of the system. 

\section{The Richardson-Gaudin Model}

\subsection{The quantum pairing hamiltonian}

Following \cite{Dukelsky}, we briefly review the Richardson pairing model.  
It describes a system of $n$ fermions characterized by a set of independent one-particle states of energies $\epsilon_l$, where the label $l$ takes values from a set $\Lambda$. The labels may refer, for instance, to orbital angular  momentum eigenstates. Each state $l$ has a total degeneracy $d_l$, and the states within the subspace corresponding to  $l$ are further labeled by an internal quantum number $s$. For instance, if the quantum number $l$ labels orbital momentum eigenstates, then $d_l = 2l+1$ and $s = -l, \ldots, l$. However, the internal degrees of freedom can be defined independently of $l$. We will assume
that $d_l$ is even, so for every state $(ls)$, there is another one related by time reversal symmetry $(l\bar s)$. For simplicity, we specialize to the case $d_l = 2, s= \uparrow, \downarrow $. Let $\hat c^{\dag}_{ls}$ represent the fermionic creation operator for the state $(ls)$. Using the Anderson pseudo-spin operators \cite{Anderson}  (quadratic pairing operators), satisfying the $su(2)$ algebra
\beq
[t^3_i, \, t_j^{\pm}] = \pm \delta_{ij} t_j^{\pm}, \,\,\,\, [t^+_i, \, t^-_j] = 2\delta_{ij}t^3_j,
\eeq
the Richardson pairing hamiltonian is given by
\beq \label{pairing}
H_P = 
\sum_{l \in \Lambda} 2\epsilon_l t^3_l - g\sum_{l, l'}t^+_l t^-_{l'} = 
\sum_{l \in \Lambda} 2\epsilon_l t^3_l - g{\bf{t}}^+\cdot {\bf{t}}^-, 
\eeq
where ${\bf {t}} = \sum_{l}{\bf {t}}_l$ is the total spin operator. 
It maps to the  reduced BCS model 
\beq \label{hamiltonian}
\hat H = \sum_{{\bf{p}}, \sigma}\epsilon_{\bf{p}}\hat{c}_{{\bf{p}}, \sigma}^{\dag}\hat{c}_{{\bf{p}}, \sigma}
-g\sum_{{\bf{p}}, {\bf{k}}}\hat{c}_{{\bf{p}} \, \uparrow}^{\dag}\hat{c}_{{-\bf{p}}\, \downarrow}^{\dag}
\hat{c}_{{-\bf{k}}\, \downarrow}\hat{c}_{{\bf{k}}\, \uparrow}
\eeq
by replacing the translational degrees of freedom by rotational ones, where $l  \in \Lambda = \{ 1, \ldots n \}$ ennumerates the one-particle orbital degrees of freedom, while $s = \, \uparrow, \, \downarrow$ indicates the two internal spin states per orbital ($d_l=2$). The pairing hamiltonian can be decomposed into the linear combination
\beq
H_P=2\sum_{l \in \Lambda} \epsilon_l R_l + g \left [ \left ( \sum_{l \in \Lambda} t^3_l \right )^2 - \frac{1}{4} \sum_{l \in \Lambda} (d_l^2-1) \right ].
\eeq
At a fixed value of the component $t^3$ of the total angular momentum, the last term becomes a constant and is dropped from the 
hamiltonian. The operators $R_l$ (generalized Gaudin magnets \cite{Gaudin76}) are given by
\beq \label{gaudin}
R_l = t^3_l - \frac{g}{2}\sum_{l' \ne l}\frac{{\bf {t}}_l \cdot {\bf {t}}_{l'}}{\epsilon_l - \epsilon_{l'}}.
\eeq
These operators solve the Richardson pairing hamiltonian because \cite{CRS97} they are independent, commute with
each other, and span all the degrees of freedom of the system. Richardson showed \cite{Richardson1,Richardson2} that the exact $N-$pair wavefunction 
of his hamiltonian is given by application of operators 
$
b^{\dag}_{k} = \sum_{l}\frac{t_l^{\dag}}{2\epsilon_l - e_k}
$
to vacuum (zero pairs state). The unnormalized $N-$pair wavefunction reads
$
\Psi_R(\epsilon_i) = \prod_{k=1}^Nb_k^{\dag}|0\rangle.
$
The eigenvalues $e_k$ satisfy the self-consistent algebraic equations 
\beq \la{ec}
\frac{1}{g} = \sum_{p \ne k}\frac{2}{e_k - e_p} +\sum_l \frac{1}{2\epsilon_l-e_k},
\eeq which can be given a 2D electrostatic interpretation \cite{Dukelsky}
with energy
\begin{eqnarray} \label{electrostatic}
U(\epsilon_l, e_k) = \frac{2}{g} \left [ \sum_{k=1}^N 
\mathcal{R}e (e_k) -
\sum_{l=1}^n \mathcal{R}e (\epsilon_l) \right ] + \\
2\sum_{l=1}^n\sum_{k=1}^N \log|e_k -2\epsilon_l| - 
4\sum_{k < p}\log|e_k-e_p|- 
\sum_{i<j}\log|2\epsilon_i-2\epsilon_j|
\end{eqnarray}
Equations (\ref{ec}) appear as equilibrium conditions for a set of charges of strength $q=2$ placed at points $e_k$, in the presence of fixed charges of strength $q=-1$ at points 
$2\epsilon_l$, and uniform electric field of strength $\frac{1}{g}$, pointing along the real axis. This interpretation proves to be very useful for the conformal field theory (CFT) description of the Richardson problem. The electrostatic energy 
(\ref{electrostatic}) is minimized for values $\{ e_k\}$ corresponding to pair energies. In (\ref{electrostatic}), $n, N$ represent the number of single-particle levels and the number of pair energies, respectively. 
For large interaction constant $g$, the equilibrium positions $\{ e_k \}$ form a set of complex conjugated pairs defining a curve $\gamma$ in the complex plane of energies. We note that the eigenvalues $r_i$ of Gaudin hamiltonia are proportional in this language to the values of the electric field at positions $2\epsilon_i$, $2r_i = g\frac{\p U}{\p \epsilon_i}. $

For a set of single-particle energies $\{\epsilon_i \}$, the BCS ground state is obtained by minimizing the electrostatic energy (\ref{electrostatic}) with respect to positions of
the free charges at $\{ e_k \}$. Once found,
they also determine exactly the values of the electric field at positions
$\{ 2\epsilon_i \}$ on the real axis, which are proportional to 
the ground-state eigenvalues  $\{ r^{GS}_i\}$. For any other values of $\{ r_i \}$, the electrostatic energy 
(\ref{electrostatic}) is not minimized. This indicates
that for arbitrary values $r_i \ne r_i^{GS}$, the system is not in equilibrium.

\subsection{The mean field limit of Richardson-Gaudin models}

\paragraph*{General description of the classical model} 
In  the mean-field limit,
the spin operators ${\bf {t}}_l$ are replaced by their quantum mechanical averages. Written in terms of the classical vectors ${\bf {S}}_l = 2 \langle {\bf {t}}_l \rangle$, the semiclassical approximation for the pairing 
hamiltonian becomes
\beq \label{meanfield}
H_{MF} = \sum_{l \in \Lambda}\epsilon_l S^3_l - \frac{g}{4}|J^-|^2,
\eeq 
where ${\bf {J}} = \sum_{l \in \Lambda} {\bf {S}}_l$ and the BCS gap function is given by $\Delta = gJ^-/2$. Replacing commutators  by canonical Poisson brackets, 
\beq \label{poisson}
\{ S^{\alpha}_i, \, S^{\beta}_j \} = 2\epsilon^{\alpha \beta \gamma}S^{\gamma}_i \delta_{ij},
\eeq
variables $S^{\alpha}_i$ become smooth functions of time. In this limit, 
the problem can analyzed with tools of classical integrable systems, and the 
solution is known to be exact as $n \to \infty$. 

The Poisson brakets (\ref{poisson}) and hamiltonian 
(\ref{meanfield}) lead to the equations of motion
\beq \label{bloch}
\dot{\vec{{S}}}_i = 2(-\vec{\Delta} + \epsilon_i \hat{z}) \times \vec{S}_i,
\eeq
where $2\vec{\Delta} = (gJ_x, gJ_y, 0)$ and $\vec{J}$ is the total spin. The 
semiclassical limit of Gaudin hamiltonia are independent constants of motion,
\beq
r_i = \frac{1}{2}\left [S_i^z - \frac{g}{2}\sum_{j \ne i}
\frac{\vec{S}_i \cdot \vec{S}_j}{\epsilon_i -\epsilon_j} \right ],
\quad
\dot r_i 
= 0.
\eeq
Equations (\ref{bloch}) describe a set of strongly interacting spins and have
generic non-linear oscillatory solutions. The exact solution may be obtained through the Abel-Jacobi inverse map \cite{Kuznetzov,Dickey,YAKE04}. 

\begin{figure} \begin{center}
 \includegraphics*[width=8cm]     {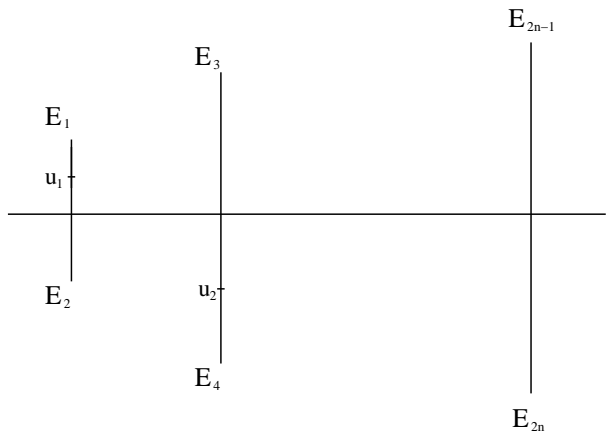}
\caption{\label{oscilat}
Schematic representation of the Liouville torus for the integrable
system (\ref{bloch}).
}
\end{center} \end{figure}

This solution can be described exactly in the language of hyperelliptic Riemann surfaces (see Appendix A for details). At this  point, it is useful to make use of the intuitively clear features of this construction (Figure~\ref{oscilat}). For a given set of initial conditions for the spins 
$\{ \vec{S}_i \}$, $i.e.$ also of the constants of motion $\{ r_i \}$, a
polynomial $Q(u)$ of degree $2n$ and with $n$ pairs of complex conjugated
roots $E_{2k+2}=\overline{E}_{2k+1}, \, k=0, \ldots, n-1$, is constructed. 
A schematic representation of these roots is given in Figure~\ref{oscilat}. 
Between each pair of roots, we place a simple cut 
$\mathcal{C}_k = [E_{2k+1}, \, E_{2k+2}]$ 
on the complex plane of energies. The surface thus obtained is a representation 
of a torus of smooth genus $g=n-1$.  

Variables $u_k$ are introduced for $n-1$ of these cuts, with respect to which the equations of motion separate. The variables $u_k$ evolve in time in a complicated fashion, solving a system of nonlinear coupled differential equations (\ref{dubrovin1}). Up to a constant, the time dependence of the gap parameter amplitude is given by 
\beq
\log |\Delta(t)| = \mathcal{I}m \, \int u(t) dt, \quad u(t) = 
\sum_k u_k (t). 
\eeq
The widths of the cuts $\mathcal{C}_k$ and the 
periods of the non-linear oscillators $u_k$ are determined by the values of
constants of motion $\{ r_i \}$. For the particular choice $r_i =r_i^{GS}$, all
the cuts $\mathcal{C}_k, \, k=1, \ldots , n-1$ vanish, and the width of the remaining cut $\mathcal{C}_n$ equals the equilibrium value of the gap function:
$|E^{GS}_{2n-1}-E^{GS}_{2n}| = 2 |\Delta|^{GS}$. In that case, the oscillators
$u_k=E_{2k-1}=E_{2k}$ are at rest, and the only time dependence left in the system is the uniform precession of the parallel planar spins $S^{-}_i$, with 
frequency $\omega = 2\sum_{k=1}^n\epsilon_k - 2\sum_{p=1}^{n-1}E_{2p-1}-\sum_{i=1}^nS^z_i$. In the case of particle-hole symmetry, $\omega$ vanishes as well. 

\subsection{The onset of pairing interaction and long-time behavior of oscillations}
Equations (\ref{bloch}) do not impose any particular 
constraints on the integrals of motion. These constants are known for the metallic state $t < 0$, but change abruptly   
during the short interval when the electron-electron interaction 
is turned on. The onset of pairing interaction is a delicate problem in itself \cite{Gurarie,Leggett,YTA05}, which deserves further investigation. There are several reasons which make the issue non-trivial. We review here several of them. 
\begin{itemize} 
\item[a)] In the case of KdV and KP2 hierarchies, it is known that finite-gap solutions (for which only a few cuts $\mathcal{C}_k$ are non-degenerate) are dense in the space of all periodic solutions. A conjecture of Krichever \cite{K1} extends this fact to arbitrary algebraic curves. Therefore, solutions where an infinite number of independent frequencies contribute to the mean-field approximation require very special initial conditions. The system is most likely to be described either by 
(i) a finite-gap solution, or (ii) a non-oscillatory function. The case (ii)
has not been investigated in previous studies. 
\item[b)] Another issue related to solutions described by only a few frequencies is 
that the $quantum$ integrability of the system may impose additional constraints and provide a selection criterion, non-existent in the mean-field approximation. The exact solution of quantum $XXX$ Gaudin system under a sudden variation of the 
spectral curve is a subject of active research and will be addressed elsewhere 
\cite{us}.  
\item[c)] Concerning the finite-gap solutions of the mean-field approximation, we note that the onset of attractive interaction is realized through some mechanism 
which effectively changes the parameters of the spectral curve. It is reasonable 
to assume that this variation is not instantaneous and in fact may continue as a
perturbation for the rest of the evolution of the system. This mechanism describes 
coupling of the system to the environment (the external magnetic field and optical trap, for degenerate Fermi gases) and therefore may induce fluctuations in the parameters describing the spectral curve. We therefore consider the effect of such 
perturbations which do not change the monodromy of the solution. The concept 
of isomonodromic deformations of integrable systems was introduced in \cite{JM1,JM2,JM3} and used extensively in the theory of Painlev\'e transcendents. We give here a brief review of the method. 
\end{itemize}

We consider the unperturbed problem given by a nonlinear scalar differential equation $R(u, \dot u, \ddot u, \ldots) = 0$, where $R$ is a rational function of the solution 
$u(t)$ and its derivatives $\dot u, \ddot u, \ldots$. Specializing 
to the cases described by hyperelliptic spectral curves, we express it through the Lax pair of $2 \times 2$ operators $L, \, A$, solving the linear vector problem
\beq \la{isospectral}
L\Psi = \mu \Psi, \,\, \p_t \Psi = A\Psi, \,\, \p_t L = [A, \, L],
\eeq  
where $\lambda, \mu$ are two auxiliary complex  variables, $A(\lambda, u)$ a $2 \times 2$ matrix, and $\Psi(\lambda, \mu, t)$ 
is a column vector. The operators are chosen such that elements of the identity $\dot L = [A, L]$ are equivalent with $R(u, \dot u, \ddot u, \ldots) = 0$. A trivial calculation gives the isospectral property $\dot \mu = 0$. 

Specializing the operator $L$ to have the form \cite{Mumford} (the $u$ functional dependence is implicit throughout) 
\beq \la{mumford}
L(\lambda) = \left [
\begin{array}{cc}
a(\lambda) & \quad b(\lambda) \\
c(\lambda) & -a(\lambda) 
\end{array}
\right ],
\eeq 
the eigenvalue equation for $\Psi$ becomes
\beq \la{crv}
\mu^2 + \det L = 0, \,\, 
\mu(\lambda) = \pm i \sqrt{\det L (\lambda)}.
\eeq 
When functions $a,b,c$ in (\ref{mumford}) are rational, equation 
(\ref{crv}) defines the hyperelliptic Riemann surface $\mu(\lambda)$
called $spectral$ $curve$ of the system (\ref{isospectral}). As noted before, the time evolution leaves the spectral curve invariant, which is why the problem is sometimes called {\it{isospectral}}.
 
The isomonodromic deformation \cite{K1} is introduced through
\beq \la{deform}
L\Psi = \mu\Psi + \epsilon \p_\lambda \Psi, \quad
[\epsilon\p_\lambda - L, \, \p_t -A] = 0,
\eeq 
where the second equation is the isomonodromy requirement, and $0 \le \epsilon \ll 1$. We recover
the unperturbed problem by setting $\epsilon = 0$. The first equation
is easily integrated and gives the formal solutions
\beq
\log \Psi_{\pm}(\lambda, \mu, t) = 
\frac{1}{\epsilon} \left [-
\mu \lambda \pm i \int ^\lambda \sqrt{\det L(\sigma)} d\sigma
\right ], 
\eeq
up to constants in $\lambda$. Let us now impose the saddle-point 
(or turning-point) condition $\p_\lambda \log \Psi = 0$. This will
give $\mu = \pm i \sqrt{\det L}$, $i.e.$ the spectral curve. Therefore, we may see the isospectral problem as a saddle-point (turning point) approximation. The compatibility equations
\beq
[\epsilon\p_\lambda - L, \, \p_t -A] = 0
\eeq
can be recast in the form 
\beq \la{split}
\epsilon(\p_\nu L - \p_\lambda A) + \p_\tau L -[A, L] = 0,
\eeq
where we have split the time dependence $\p_t$ into a fast time scale
$\p_\tau$ and a slow one $\epsilon \p_\nu$. At zero order  in $\epsilon$, (\ref{split}) is simply the unperturbed problem. The first order correction gives the slow-time scale dependence of the modulated solution $u_\epsilon(\tau, \nu)$. Moreover, from the matrix elements of (\ref{split}), we get after averaging over the fast motion in $\tau$ \cite{Takasaki},
\beq \la{det}
\p_\nu \det L = -(\overline{2aA_{11}+bA_{21}+cA_{12}}),
\eeq
where the bar signifies $\tau-$averaging. Since for the unperturbed problem, $\p_\tau \det L = 0$, the averaging is justified.

Equation (\ref{det}) is simply the result of Bogoliubov-Whitham \cite{Whitham} averaging for the problem (\ref{deform}). It tells us how the previously invariant spectral curve now changes slowly in time over the large time scale $\nu$. It also gives us the form of the
deformed solution $u_{\epsilon}(\tau, k_i(\nu))$, as a {\it{modulation}} of the original solution.  

\paragraph*{Note} The fact that such perturbations generalize the autonomous  Garnier system of \cite{YAKE04} to a non-autonomous system of Schlesinger type was indicated in \cite{Takasaki}. Extensions of these systems to include a constant matrix were discussed in \cite{Beauville}. .

\section{The effect of weak perturbations}

\subsection{Topological classification of finite-gap solutions} \la{whith}
 
In the presence of spectrum symmetry $\epsilon_k = - \epsilon_{-k}, S_k^z = -S_{-k}^z$, the 
distribution of cuts $\mathcal{C}_k$ obeys the same symmetry. The number of non-degenerate 
cuts is therefore even. This analysis uses the fact that at $t=0$ all Anderson pseudo-spins are aligned along the $z$ axis. In the simplest non-trivial case, there are only two non-degenerate cuts as 
shown in Figure~\ref{genus1} and $g=1$, while the corresponding variable $u_1$ is given by elliptic functions. Other non-trivial cuts $\mathcal{C}_k$ may exist in general, associated with dynamics of variables $u_k$. In the limit of ``small" cuts \cite{Bobenko}, their contributions 
separate are given by trigonometric functions and the behavior of the gap parameter takes the 
simplified form
\beq \la{expansion}
\log \frac{|\Delta(t)|}{|\Delta(0)|} = \mathcal{I}m \, \int u(t) dt = \mathcal{U}_{ell}(t) + \sum_k
\mathcal{U}^k_{trig}.
\eeq
As we shall see, the most relevant contribution is due to the elliptic part $\mathcal{U}_{ell}$, studied in the next section.

\subsection{Modulations of elliptic solutions}

\begin{figure} \begin{center}
 \includegraphics*[width=8cm]     {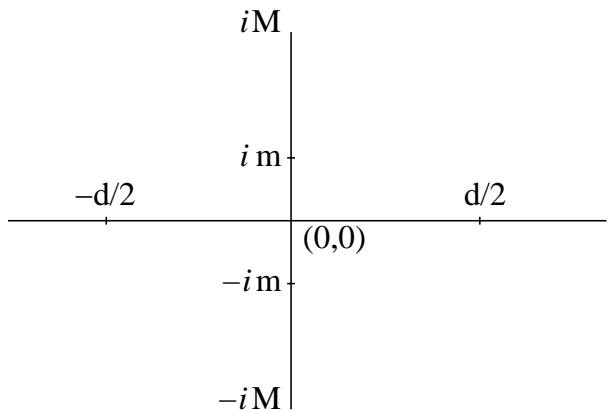}
\caption{\label{genus1}
Distribution of complex roots $E_i$ around the origin, 
for the elliptic approximation. The mean level spacing is $d$.}
\end{center} \end{figure}

For the root distribution shown in Figure~\ref{genus1}, there is only one
variable $u_k$, taking imaginary values \cite{YAKE04}. It 
solves an equation of the type \cite{YAKE04}
\beq
(\dot u)^2 + (u^2 + m^2)(u^2 + M^2) = 0.
\eeq
For simplicity, we therefore make the transformation $u \to iu$, keeping the time variable real. The new function is a real quartic oscillator which satisfies
\beq \la{osc}
(\dot u)^2 = (u^2 - M^2)(u^2 - m^2).
\eeq 
The solutions corresponding to this distribution of roots are 
\begin{eqnarray} \label{u1}
u_1(t) = m\cdot sn (Mt + \phi_1, k_1), \quad
k_1 = m/M, \\
u_2(t) = M\cdot sn (mt + \phi_2, k_2), \quad \,
k_2 = M/m, 
\end{eqnarray}
where $sn$ is the Jacobi sine function, and $\phi_{1,2}$ 
are arbitrary phases. In the degenerate case $m=M$, the 
solutions become hyperbolic functions. Solution $u_2$ is non-physical in our case.

In order to set up the isomonodromic deformation method, consider the Lax pair \cite{Its}:
\begin{eqnarray}
\la{L1}
L = -\left [ 
2\dot u \sigma_1 +
4u\lambda \sigma_2 +
(4\lambda^2 - \xi  + 2u^2)i\sigma_3
\right ], \\
\la{L2}
A = i\lambda \sigma_3 - u\sigma_2,
\,\,\,\,\,\,\,
\epsilon \p_\lambda \Psi = L \Psi, \,\,\,\,\,\,\,
\p_t  \Psi = A \Psi, 
\end{eqnarray}
where $\sigma_\alpha$, $\alpha = 1, 2, 3$ are the Pauli 
matrices, $\epsilon$ is a small real number, 
and $\Psi = (\psi_1, \, \psi_2)^t$ is the Baker-Akhiezer
function. Choosing the initial condition 
$\psi_1 = 0, \, \psi_2 = 1$, we can extract the amplitude of the ratio $\Delta(t)/\Delta(0)$ from $\Psi(t)$ as
\beq
\left | \frac{\Delta(t)}{\Delta(0)} \right | = \frac{
\left [ \psi_1(t) + \overline{\psi}_1(t)\right ]
}{2}.
\eeq
The compatibility (zero-curvature) conditions
\beq \label{compat}
[\p_t - A , \,  \epsilon \p_\lambda - L] = 0
\eeq
yield the system of equations
\beq
\p_t \xi = \epsilon, \quad \p_t^2 u = 2u^3-\xi u.
\eeq 
We note that
\beq
\det L = -4 \{ (\dot u)^2 - [ u^4 - \xi u^2 + 
(2\lambda^2 - \xi/2)^2] \}.  
\eeq
Setting $\epsilon = 0$ gives the equation $(\dot u)^2 - u^4 + \xi u^2 = $ constant, while the limit $\epsilon = 1$ yields the Painlev\'e II equation. In fact, the unperturbed case $\epsilon = 0$ allows
to retrieve the full elliptic solution, from the equation
\beq
L\Psi =0, \quad \det(L) = 0,
\eeq  
which gives the elliptical function $u$ satisfying
\beq
(\dot u)^2 - [ u^4 - \xi u^2 + (2\lambda^2 - \xi/2)^2] = 0. 
\eeq
The physical solution $u_1$ is obtained identifying
\beq
\xi = m^2+M^2, \quad 2 \lambda = M + m.  
\eeq
Now restore $\epsilon \ne 0$, write $\p_t = \p_{\tau} 
+\epsilon \p_{\xi}$, and retain terms of order $\epsilon$
from (\ref{compat}). Averaging over the fast variable $\tau$ 
gives
\beq
\p_\xi \det L = 
\overline{L_{22}\p_\lambda A_{11}} +
\overline{L_{11}\p_\lambda A_{22}}
,
\eeq
or equivalently,
\beq
\p_{\xi} \det L 
= -(4\overline{u^2} - 2\xi +8\lambda^2).
\eeq
Performing the computations, we obtain
\beq
\p_\xi [u^4 -\xi u^2 -(\dot u)^2] = - \overline{u^2}.   
\eeq
Writing the elliptic equation as
\beq \la{c1}
(\dot u)^2 = u^4 -\xi u^2 + \mu^2,
\eeq
the physical solution takes the form
$$
u_1(t) = \sqrt{\frac{\xi - \sqrt{\xi^2 - 4\mu^2}}{2}}
sn \left [\sqrt{\frac{\xi + \sqrt{\xi^2 - 4\mu^2}}{2}}t \right ],
$$
up to an arbitrary initial phase $\phi$, and elliptic modulus
\beq
k^2 = \frac{1 - \sqrt{1-(\frac{2\mu}{\xi})^2}}
{1 + \sqrt{1-(\frac{2\mu}{\xi})^2}}.
\eeq
The Whitham averaging equation has the form
\beq \label{whitham}
4\frac{\p \mu^2}{\p \xi^2} =1-
\frac{2}{2-k^2}\frac{\mathcal{E}(k^2)}{\mathcal{K}(k^2)},
\eeq
where $\mathcal{E, K}$ are the complete elliptic integrals of the
first and second kind, respectively. 
Equation (\ref{whitham}) has a fixed point at $k = 0$, $\frac{\mu}{\xi} \to 0$. This shows that, on the slow time scale,
the parameter $\mu/\xi = m/M + O(m^3/M^3)$ goes to zero, as $\xi = m^2 + M^2$ increases. Expanding the solution 
in this limit, and integrating under the separation of time scales 
asumption, we obtain for the elliptic contribution to the gap parameter, the approximation
\beq
\Delta_{ell}(t) = \Delta(0) e^{k \int sn(\tau, k^2 ) d\tau},
\eeq
and $k^2 = \frac{\mu^2}{\xi^2} + O(\mu^4/\xi^4) \to 0$ as $t \to \infty$. 

\paragraph*{Asymptotic behavior of modulated elliptic solutions}
Starting from the Lax pair (\ref{L1}, \ref{L2}), we can obtain 
the asymptotic behavior of $\Psi(\lambda, t)$ as $t \to \infty$, 
in the whole complex $\lambda$ plane \cite{Its}. The analysis is
simpler when working in the variable $z = \frac{\lambda}{\sqrt{\xi}}$. There are six Stokes sectors at $z \to \infty$, with canonical
asymptotes for $\Psi$, but the region of interest is $z = 1/2$, 
where 
\beq
\frac{4\lambda^2}{\xi} \to 1, \quad \frac{\mu^2}{\xi^2} \to 0.
\eeq
Using the Whitham equation, separation of scales, and identifying the strength of the fluctuations $\epsilon$ with an effective temperature $T$, we obtain for the elliptic contribution 
at large times (\ref{expansion})
\beq
U_{ell}(t) = \frac{\mathcal{F}(t)}{tT},
\eeq
where $\mathcal{F}$ is a bounded oscillatory function 
$|\mathcal{F}| = O(1)$. 

\paragraph*{Modulations of trigonometric solutions}
In order to analyze the slow dynamics of the small cuts, 
let $\xi \to \infty$ in (\ref{c1}) and write the solution as
\beq \la{trg}
u = \frac{\mu}{\sqrt{\xi}} \cos(\sqrt{\xi}t + \phi), 
\quad \mu^2/\xi \to 0.
\eeq
This assumption is consistent with (\ref{c1}). Formally, this is simply the trigonometric 
approximation of the general elliptic solution, in
the limit of small modulus $k^2$. However, the parameter $\xi$ in (\ref{trg}) must be sent to $\infty$ mush faster than the physical $\xi = M^2 + m^2$, in order to give the correct spherical limit.
 The contributions from the small cuts therefore vanish faster
than the elliptic component.

\paragraph*{Higher-genus contributions}
In the case where there are several non-degenerate cuts at $t=0$ corresponding to a high 
genus $g>1$, the isomodromic deformation method can be applied in the same way as for $g=1$, 
leading to modulated hyper-elliptic solutions of Painlev\'e equations degree higher than 2. There
are few systematic results in this field, and a complete classification does not exist at this time. 
Specific high-genus solutions may have interesting topological properties with physical 
interpretations in terms of the collective excitations of the Gaudin magnet \cite{us}.

\appendix{} \la{Lax}

\section{Isospectral case and the Abel-Jacobi inversion problem}

In \cite{Kuznetzov,YAKE04}, the system (\ref{bloch}) with fixed spectral curve was solved through inverse Abel-Jacobi mapping, by using Sklyanin separation of variables techniques \cite{SOV,Smirnov}. 
Interesting connections to generalized Neumann systems and Hitchin systems were discovered in \cite{Hikami}.
The solution starts from the Lax operator 
\beq \label{lax} 
\mathcal{L}(\lambda) = \frac{2}{g}\sigma_3 +\sum_{i=0}^n \frac{\vec S_i \cdot\vec \sigma}{\lambda - \epsilon_i} = 
\left [
\begin{array}{cr}
a(\lambda) & \,\, \, b(\lambda) \\
c(\lambda) & -a(\lambda)
\end{array}
\right ],
\eeq
where $\sigma_\alpha, \, \alpha = 1,2,3$ are the Pauli matrices, and $\lambda$ is an additional complex variable, the spectral 
parameter. Let $u_k, \, k=1, \ldots, n-1$ be the roots of the coefficient $c(\lambda)$. 
Poisson brackets for variables $S_i^{\alpha}$ read 
\beq
\{S_j^\alpha , \, S_k^{\beta} \}=2\epsilon_{\alpha \beta \gamma}S_k^\gamma \delta_{jk}
\eeq
The Lax operator (\ref{lax}) defines 
a Riemann surface (the spectral curve) $\Gamma (y, \lambda)$ of genus 
$g = n-1$, through 
\beq \label{curve}
y^2 = Q(\lambda) = \det \mathcal{L}(\lambda)\left [ g\frac{P(\lambda)}{2} \right ]^2 , 
\eeq
where $P(\lambda) = \prod_{i=1}^n (\lambda -  \epsilon_i).$

The equations of motion for the hamiltonian (\ref{meanfield}) become
\beq \label{dubrovin1}
\dot  u_i = \frac{2i y( u_i)}{\prod_{j \ne i}( u_i - u_j)},
\quad
i\dot J^- = J^{-} \left [ gJ^3 + 2\sum_{k=1}^{n} \epsilon_k - u \right ].
\eeq
In (\ref{dubrovin1}), 
$u = -2\sum_{i=1}^{n-1}  u_i, \quad b(u_i) = 0$. 

From the equations of motion, it is clear that knowledge of the initial amplitude of $J^-$ and of the roots $ \{  u_i \}$
is enough to specify the $n$ unit vectors $\{ {\bf {S}}_i \},$ for a given set of constants of motion $\{ R_l \}$ given by the classical limit of Gaudin hamiltonia. The Dubrovin equations (\ref{dubrovin1}) are 
solved by the inverse of the Abel-Jacobi map, as we explain in the following. We begin by noting that the polynomial 
$Q(\lambda)$ has degree $2n$, and is positively defined on the real $\lambda$ axis. Therefore, the curve $\Gamma (y, \lambda)$
has $n$ cuts between the pairs of complex roots $[E_{2i-1}, \, E_{2i}], i = 1, 2, \ldots, n$, perpendicular to the real $\lambda$
axis. The points $u_i$ belong to $n-1$ of these cuts, $u_i \in [E_{2i-1}, \, E_{2i}], i =1, \ldots, n-1$. These 
$g = n-1$ cuts allow to define a canonical homology basis of $\Gamma$, consisting of cycles $\{\alpha_i, \beta_i \}, i = 1, \ldots, g$. With respect to these cycles, a basis of normalized holomorphic differentials $\{ \omega_i \}$ can be defined, through
\beq
\mu_i = \lambda^{g-i}\frac{d\lambda}{y}, \,\,\, M_{ij} = \int_{\alpha_j} \mu_i, \,\,\, {\bm {\omega}} = M^{-1}{\bm{\mu}}. 
\eeq 
The period matrix $B_{ij} = \int_{\beta_j} \omega_i$
is symmetric and has positively defined imaginary part. The Riemann $\theta$ function is defined with the help of the period matrix as
\beq
\theta({\bm{z}} | B) = \sum_{{\bm{n}} \in {\bm{z}}^g} e^{2\pi i ({\bm {n}}^t {\bm{z}} + \frac{1}{2} {\bm {n}}^t B {\bm {n}} )}.
\eeq
The $g$ vectors $\bm{B}_k$ consisting of columns of $B$ and the basic
vectors $\bm{e}_k$  define a lattice in ${\mathbb{C}}^g$. The $Jacobian$ variety of the curve $\Gamma$, is then the $g-$dimensional torus defined as the quotien 
$J(\Gamma)  = \mathbb{C}^g /(\mathbb{Z}^g + B\mathbb{Z}^g).$
The Abel-Jacobi map associates to any point $P$ on $\Gamma$, a point ($g-$ dimensional complex vector) on the Jacobian variety, through
${\bm {A}} (P) = \int_{\infty}^P {\bm {\omega}}.$
Considering now a $g-$dimensional complex vector of points $\{P_k \}, 
k = 1, \ldots, g$ on $\Gamma$, defined up to a permutation, we can associate to it the point on the Jacobian
\beq  \label{a}
{\bm{z}} = {\bm{a}}({\bm {P}}) = \sum_{k=1}^g {\bm {A}} (P_k) + {\bm {K}},
\eeq
where ${\bm {K}}$ is the Riemann characteristic vector for $\Gamma$. 

The map (\ref{a}) suggests that we now have a 
way to describe the dynamics on $\Gamma$ by following the image point on the Jacobian. 
Given a point on the $g-$dimensional Jacobian  ${\bm{z}} = (\zeta_1, \ldots , \zeta_{n-1})$, we can find an unique set of points $\{ \lambda_k\}, k = 1, \ldots, g$ on $\Gamma$, such that ${\bm{z}} = {\bm {a}}(
{\bm {\lambda}}),$ and $\theta ({\bm{a}(\bm{P}) - \bm{z}} | B) = 0$. The system evolves in time according to the point ${\bm{z}}(t)$ 
\beq \label{solution1}
\zeta_k = ic_k, \,\, 1 \le i \le g-1, \,\, \zeta_{n-1} = i(c_{n-1} + t),
\eeq
where $\{ c_k\}$ is a set of initial conditions, such that  $
{\bm{z}}_0 = {\bm{z}}(t=0) = {\bm {a}}({\bm{c}})$, and ${\bm{c}}$ is the set of initial
conditions for positions of ${\bm {\lambda}}$ on $\Gamma$. Together with the initial condition which determines the
initial amplitude of $J^-$, this set will determine entirely the evolution of the functions $ u_i(t), J^-(t)$.

\subsection*{Acknowledgments}

The author is grateful to I~Aleiner, I~Gruzberg, I~Krichever and P~Wiegmann for suggestions and contributions. Useful discussions with A~G~Abanov, B~Altshuler, E~Bettelheim, and E~Yuzbashyan are acknowledged. The author also thanks I~Aleiner and A~Millis for support.

\subsection*{References}

\bibliography{references}

\begin{thebibliography}{10}

\bibitem{Richardson1}
R.W. Richardson.
\newblock {\em Phys. Lett.}, 3:277, 1963.

\bibitem{Richardson2}
R.W. Richardson.
\newblock {\em Phys. Lett.}, 5:82, 1963.

\bibitem{Richardson3}
R.~W. Richardson and N.~Sherman.
\newblock Exact eigenstates of the pairing-force {H}amiltonian.
\newblock {\em Nuclear Phys.}, 52:221--238, 1964.

\bibitem{Richardson4}
R.~W. Richardson.
\newblock Exact eigenstates of the pairing-force {H}amiltonian. {II}.
\newblock {\em J. Mathematical Phys.}, 6:1034--1051, 1965.

\bibitem{Richardson5}
R.W. Richardson.
\newblock {\em Phys. Rev.}, 141:949, 1966.

\bibitem{Richardson6}
R.W. Richardson.
\newblock {\em J. Math. Phys.}, 9:1327, 1968.

\bibitem{Gaudin76}
M.~Gaudin.
\newblock Diagonalisation d'une classe d'{H}amiltoniens de spin.
\newblock {\em J. Physique}, 37(10):1089--1098, 1976.

\bibitem{SOV}
E.~K. Sklyanin.
\newblock Separation of variables in the {G}audin model.
\newblock {\em Zap. Nauchn. Sem. Leningrad. Otdel. Mat. Inst. Steklov. (LOMI)},
  164(Differentsialnaya Geom. Gruppy Li i Mekh. IX):151--169, 198, 1987.

\bibitem{Kuznetzov}
A.~N.~W. Hone, V.~B. Kuznetsov, and O.~Ragnisco.
\newblock B\"acklund transformations for the {${\rm sl}(2)$} {G}audin magnet.
\newblock {\em J. Phys. A}, 34(11):2477--2490, 2001.

\bibitem{Kuz1}
V.~B. Kuznetsov.
\newblock Quadrics on {R}iemannian spaces of constant curvature. {S}eparation
  of variables and a connection with the {G}audin magnet.
\newblock {\em Teoret. Mat. Fiz.}, 91(1):83--111, 1992.

\bibitem{Kuz2}
V.~Kuznetsov and P.~Vanhaecke.
\newblock B\"acklund transformations for finite-dimensional integrable systems:
  a geometric approach.
\newblock {\em J. Geom. Phys.}, 44(1):1--40, 2002.

\bibitem{Kuz3}
E.~G. Kalnins, V.~B. Kuznetsov, and Willard Miller, Jr.
\newblock Quadrics on complex {R}iemannian spaces of constant curvature,
  separation of variables, and the {G}audin magnet.
\newblock {\em J. Math. Phys.}, 35(4):1710--1731, 1994.

\bibitem{Kuz4}
V.~B. Kuznetsov.
\newblock Isomorphism of an {$n$}-dimensional {N}eumann system and an
  {$n$}-site {G}audin magnet.
\newblock {\em Funktsional. Anal. i Prilozhen.}, 26(4):88--90, 1992.

\bibitem{Kuz5}
V.~B. Kuznetsov.
\newblock Equivalence of two graphical calculi.
\newblock {\em J. Phys. A}, 25(22):6005--6026, 1992.

\bibitem{YAKE04}
E.~A. {Yuzbashyan}, B.~L. {Altshuler}, V.~B. {Kuznetsov}, and V.~Z. {Enolskii}.
\newblock {Nonequilibrium Cooper Pairing in the Non-adiabatic Regime}.
\newblock {\em ArXiv Condensed Matter e-prints}, May 2005.

\bibitem{Sierra}
G.~Sierra.
\newblock Integrability and conformal symmetry in the {BCS} model.
\newblock In {\em Statistical field theories (Como, 2001)}, volume~73 of {\em
  NATO Sci. Ser. II Math. Phys. Chem.}, pages 317--328. Kluwer Acad. Publ.,
  Dordrecht, 2002.

\bibitem{Levitov}
R.~A. {Barankov}, L.~S. {Levitov}, and B.~Z. {Spivak}.
\newblock {Collective Rabi Oscillations and Solitons in a Time-Dependent BCS
  Pairing Problem}.
\newblock {\em Phys. Rev. Lett.}, 93(16):160401, October 2004.

\bibitem{CRS97}
M.~C. {Cambiaggio}, A.~M.~F. {Rivas}, and M.~{Saraceno}.
\newblock {Integrability of the pairing hamiltonian}.
\newblock {\em Nuclear Physics A}, 624:157--167, February 1997.

\bibitem{ALO21}
L.~{Amico}, A.~{di Lorenzo}, and A.~{Osterloh}.
\newblock {Integrable Model for Interacting Electrons in Metallic Grains}.
\newblock {\em Phys. Rev. Lett.}, 86:5759--5762, June 2001.

\bibitem{DES01}
J.~{Dukelsky}, C.~{Esebbag}, and P.~{Schuck}.
\newblock {Class of Exactly Solvable Pairing Models}.
\newblock {\em Phys. Rev. Lett.}, 87(6):066403, August 2001.

\bibitem{Dukelsky}
J.~Dukelsky, S.~Pittel, and G.~Sierra.
\newblock Colloquium: {E}xactly solvable {R}ichardson-{G}audin models for
  many-body quantum systems.
\newblock {\em Rev. Modern Phys.}, 76(3):643--662, 2004.

\bibitem{AFS}
M.~Asorey, F.~Falceto, and G.~Sierra.
\newblock Chern-{S}imons theory and {BCS} superconductivity.
\newblock {\em Nuclear Phys. B}, 622(3):593--614, 2002.

\bibitem{Samtleben}
N.~Manojlovi{\'c} and H.~Samtleben.
\newblock Schlesinger transformations and quantum {$R$}-matrices.
\newblock {\em Comm. Math. Phys.}, 230(3):517--537, 2002.

\bibitem{Hikami}
K.~Hikami.
\newblock Separation of variables in the {BC}-type {G}audin magnet.
\newblock {\em J. Phys. A}, 28(14):4053--4061, 1995.

\bibitem{Gawedzki1}
F.~Falceto and K.~Gaw{\c{e}}dzki.
\newblock Unitarity of the {K}nizhnik-{Z}amolodchikov-{B}ernard connection and
  the {B}ethe ansatz for the elliptic {H}itchin systems.
\newblock {\em Comm. Math. Phys.}, 183(2):267--290, 1997.

\bibitem{Gawedzki2}
F.~Falceto and K.~Gaw{\c{e}}dzki.
\newblock Chern-{S}imons states at genus one.
\newblock {\em Comm. Math. Phys.}, 159(3):549--579, 1994.

\bibitem{Ortiz}
G.~Ortiz, R.~Somma, J.~Dukelsky, and S.~Rombouts.
\newblock Exactly-solvable models derived from a generalized {G}audin algebra.
\newblock {\em Nuclear Phys. B}, 707(3):421--457, 2005.

\bibitem{Korotkin}
D.~Korotkin and H.~Samtleben.
\newblock On the quantization of isomonodromic deformations on the torus.
\newblock {\em Internat. J. Modern Phys. A}, 12(11):2013--2029, 1997.

\bibitem{Harnad}
J.~Harnad.
\newblock Quantum isomonodromic deformations and the {K}nizhnik-{Z}amolodchikov
  equations.
\newblock In {\em Symmetries and integrability of difference equations
  (Est\'erel, PQ, 1994)}, volume~9 of {\em CRM Proc. Lecture Notes}, pages
  155--161. Amer. Math. Soc., Providence, RI, 1996.

\bibitem{Dilorenzo}
A.~Di~Lorenzo, L.~Amico, K.~Hikami, A.~Osterloh, and G.~Giaquinta.
\newblock Quasi-classical descendants of disordered vertex models with
  boundaries.
\newblock {\em Nuclear Phys. B}, 644(3):409--432, 2002.

\bibitem{Fermi1}
B.~{Demarco} and D.~S. {Jin}.
\newblock {\em Science}, 285:1703, 1999.

\bibitem{Fermi2}
B.~{Demarco}, J.~L. {Bohn}, J.~P. {Burke}, M.~{Holland}, and D.~S. {Jin}.
\newblock {Measurement of p-Wave Threshold Law Using Evaporatively Cooled
  Fermionic Atoms}.
\newblock {\em Phys. Rev. Lett.}, 82:4208--4211, May 1999.

\bibitem{Fermi3}
B.~{Demarco}, S.~B. {Papp}, and D.~S. {Jin}.
\newblock {Pauli Blocking of Collisions in a Quantum Degenerate Atomic Fermi
  Gas}.
\newblock {\em Phys. Rev. Lett.}, 86:5409--5412, June 2001.

\bibitem{Fermi4}
B.~{Demarco} and D.~S. {Jin}.
\newblock {Spin Excitations in a Fermi Gas of Atoms}.
\newblock {\em Phys. Rev. Lett.}, 88(4):040405, January 2002.

\bibitem{Fermi5}
A.~{Truscott}, K.~{Strecker}, G.~{Partridge}, R.~{Hulet}, and R.~{Hulet}.
\newblock {\em Science}, 291:2570, 2001.

\bibitem{Fermi6}
T.~{Loftus}, C.~A. {Regal}, C.~{Ticknor}, J.~L. {Bohn}, and D.~S. {Jin}.
\newblock {Resonant Control of Elastic Collisions in an Optically Trapped Fermi
  Gas of Atoms}.
\newblock {\em Phys. Rev. Lett.}, 88(17):173201, April 2002.

\bibitem{Fermi7}
K.~M. {O'Hara}, S.~L. {Hemmer}, M.~E. {Gehm}, S.~R. {Granade}, and J.~E.
  {Thomas}.
\newblock {Observation of a Strongly Interacting Degenerate Fermi Gas of
  Atoms}.
\newblock {\em Science}, 298:2179--2182, December 2002.

\bibitem{Fermi8}
K.~M. {O'Hara}, S.~L. {Hemmer}, S.~R. {Granade}, M.~E. {Gehm}, J.~E. {Thomas},
  V.~{Venturi}, E.~{Tiesinga}, and C.~J. {Williams}.
\newblock {Measurement of the zero crossing in a Feshbach resonance of
  fermionic Li}.
\newblock {\em Phys. Rev. A}, 66(4):041401, October 2002.

\bibitem{Anderson}
P.~W. {Anderson}.
\newblock {Random-Phase Approximation in the Theory of Superconductivity}.
\newblock {\em Physical Review}, 112:1900--1916, December 1958.

\bibitem{Gurarie}
A.~V. {Andreev}, V.~{Gurarie}, and L.~{Radzihovsky}.
\newblock {Nonequilibrium Dynamics and Thermodynamics of a Degenerate Fermi Gas
  Across a Feshbach Resonance}.
\newblock {\em Phys. Rev. Lett.}, 93(13):130402, September 2004.

\bibitem{Leggett}
G.~L. {Warner} and A.~J. {Leggett}.
\newblock {Quench dynamics of a superfluid Fermi gas}.
\newblock {\em Phys. Rev. B}, 71(13):134514, April 2005.

\bibitem{YTA05}
E.~A. {Yuzbashyan}, O.~{Tsyplyatyev}, and B.~L. {Altshuler}.
\newblock {Relaxation and persistent oscillations of the order parameter in the
  non-stationary BCS theory}.
\newblock {\em ArXiv Condensed Matter e-prints}, November 2005.

\bibitem{Krichever}
I.~Krichever.
\newblock Isomonodromy equations on algebraic curves, canonical transformations
  and {W}hitham equations.
\newblock {\em Mosc. Math. J.}, 2(4):717--752, 806, 2002.
\newblock Dedicated to Yuri I. Manin on the occasion of his 65th birthday.

\bibitem{K1}
I.~Krichever.
\newblock Vector bundles and {L}ax equations on algebraic curves.
\newblock {\em Comm. Math. Phys.}, 229(2):229--269, 2002.

\bibitem{Dickey}
L.~A. Dickey.
\newblock {\em Soliton equations and {H}amiltonian systems}, volume~26 of {\em
  Advanced Series in Mathematical Physics}.
\newblock World Scientific Publishing Co. Inc., River Edge, NJ, 2003.

\bibitem{us}
I.~Aleiner and R.~Teodorescu.
\newblock Unpublished.

\bibitem{JM1}
M.~Jimbo and T.~Miwa.
\newblock Monodromy preserving deformation of linear ordinary differential
  equations with rational coefficients. {III}.
\newblock {\em Phys. D}, 4(1):26--46, 1981/82.

\bibitem{JM2}
M.~Jimbo, T.~Miwa, and K.~Ueno.
\newblock Monodromy preserving deformation of linear ordinary differential
  equations with rational coefficients. {I}. {G}eneral theory and {$\tau
  $}-function.
\newblock {\em Phys. D}, 2(2):306--352, 1981.

\bibitem{JM3}
M.~Jimbo and T.~Miwa.
\newblock Monodromy preserving deformation of linear ordinary differential
  equations with rational coefficients. {II}.
\newblock {\em Phys. D}, 2(3):407--448, 1981.

\bibitem{Mumford}
D.~Mumford.
\newblock {\em Tata lectures on theta. {II}}, volume~43 of {\em Progress in
  Mathematics}.
\newblock Birkh\"auser Boston Inc., Boston, MA, 1984.

\bibitem{Takasaki}
K.~Takasaki.
\newblock Spectral curves and {W}hitham equations in isomonodromic problems of
  {S}chlesinger type.
\newblock {\em Asian J. Math.}, 2(4):1049--1078, 1998.

\bibitem{Whitham}
G.~B. Whitham.
\newblock {\em Linear and nonlinear waves}.
\newblock Pure and Applied Mathematics (New York). John Wiley \& Sons Inc., New
  York, 1999.

\bibitem{Beauville}
A.~Beauville.
\newblock Jacobiennes des courbes spectrales et syst\`emes hamiltoniens
  compl\`etement int\'egrables.
\newblock {\em Acta Math.}, 164(3-4):211--235, 1990.

\bibitem{Bobenko}
E.D. Belokolos, A.I. Bobenko, V.Z. Enol'skii, A.R. Its, and V.B. Matveev.
\newblock {\em Algebro-Geometric Approach to Nonlinear Integrable Equations},
  volume~11 of {\em Springer Series in Nonlinear Dynamics}.
\newblock Springer-Verlag, Berlin, 1994.

\bibitem{Its}
A.~R. Its.
\newblock The {P}ainlev\'e transcendents as nonlinear special functions.
\newblock In {\em Painlev\'e transcendents (Sainte-Ad\`ele, PQ, 1990)}, volume
  278 of {\em NATO Adv. Sci. Inst. Ser. B Phys.}, pages 49--59. Plenum, New
  York, 1992.

\bibitem{Smirnov}
A.~Nakayashiki and F.~A. Smirnov.
\newblock Cohomologies of affine hyperelliptic {J}acobi varieties and
  integrable systems.
\newblock {\em Comm. Math. Phys.}, 217(3):623--652, 2001.

\end{thebibliography}
\bibliographystyle{unsrt}

\end{document}